\documentstyle[12pt,epsf]{article}
\begin{document}
\newcommand {\ecm} {E$_{\,\mathrm{c.m.}}$}
\newcommand {\ecmm} {\mathrm{E}_{\,\mathrm{c.m.}}}
\newcommand {\epem}     {e$^+$e$^-$}
\newcommand {\mnch} {$\langle n_{\,\mathrm{ch.}} \rangle$}
\newcommand {\gluglu} {gg}
\newcommand {\qbar} {$\overline{\mathrm{q}}$}
\newcommand {\qq} {q$\overline{\mathrm{q}}$}
\newcommand {\gincl} {g$_{\,\mathrm{incl.}}$}
\newcommand {\ejet} {$E_{\,\mathrm{jet}}$}
\newcommand {\durham}   {$k_\perp$}
\newcommand {\ngprime} {n_{\mathrm{G}}^{\prime}}
\newcommand {\ng} {n_{\mathrm{G}}}
\newcommand {\nf} {n_{\mathrm{F}}}
\newcommand {\ratio} {r}
\newcommand {\rprime} {r^{\prime}}
\newcommand {\qzero} {\mathrm{Q}_0}
\newcommand {\aone} {a_1}
\newcommand {\atwo} {a_2}
\newcommand {\athree} {a_3}
\newcommand {\afour} {a_4}
\newcommand {\rzero} {r_0}
\newcommand {\rone} {r_1}
\newcommand {\rtwo} {r_2}
\newcommand {\rthree} {r_3}
\newcommand {\ncc} {\mathrm{N}_{\mathrm{C}}}
\newcommand {\nff} {\mathrm{N}_{\mathrm{F}}}
\newcommand {\caa} {\mathrm{C}_{\mathrm{A}}}
\newcommand {\cff} {\mathrm{C}_{\mathrm{F}}}
\newcommand {\cc} {\mathrm{C}}

\pagestyle{empty}
\vspace*{-1cm}
\begin{flushright}
  FIAN-30/99 \\
  UCRHEP-E255 \\
  14 May 1999 \\
  revised 24 August 1999
\end{flushright}

\vspace*{1cm}
\baselineskip 14pt
\begin{center}{\LARGE\bf 
Energy dependence of mean multiplicities in
gluon and quark jets at the 
next-to-next-to-next-to-leading order
}\end{center}\bigskip
\begin{center}
\large I.M. Dremin$^{1}$ and J.W. Gary$^{2}$
\end{center}
\begin{center}
{\it $^{1}$Lebedev Physical Institute, Moscow 117924, Russia\\
$^{2}$Department of Physics, University of California, 
  Riverside CA 92521, USA}
\end{center}

\vspace*{.2cm}
\begin{abstract}
Analytic predictions for the energy dependence of the
mean multiplicities in gluon and quark jets are presented at
the next-to-next-to-next-to-leading order
(3NLO) of perturbative QCD
and are compared to experiment.
The 3NLO correction to the gluon jet multiplicity is
found to be small.
The corresponding
theoretical expression provides a good
description of available gluon jet measurements.
The 3NLO correction to the quark jet multiplicity 
is large.
The theoretical expression for quark jets also
describes the data accurately, however,
but not with the same parameter values as are
used for gluon jets.
It is shown that the well known
success of the next-to-leading order (NLO) 
approximation in describing the energy evolution of
quark jet multiplicity can be attributed to the
equivalence of the quark and gluon expressions at NLO
to within a constant factor,
and to almost constant contributions from higher
order terms to the gluon jet result.
\
\end{abstract}

\vspace*{1cm}
\noindent
pacs: 12.38.Bx,12.38.-t, 13.65.+i, 14.70.Dj \\
keywords:  Jets, Multiplicity, Perturbative QCD, Gluons

\newpage
\pagenumbering{arabic}
\pagestyle{plain}

Particle multiplicity is one of the most basic 
properties of a jet.
The multiplicity of a jet
has been the subject of many theoretical
and experimental studies~\cite{bib-reviews}.
Analytic predictions for
the energy dependence of mean
multiplicities in gluon and quark jets
can be derived from perturbative QCD (pQCD).
The total multiplicity of a jet,
i.e.~the multiplicity in full phase space,
is subject to important corrections in pQCD
and thus provides a sensitive means to test
higher order perturbation theory.\footnote{This 
is in contrast to multiplicity in a limited 
region of phase space,
such as the multiplicity of soft particles,
which is less sensitive to 
higher order corrections;
this fact has recently
been exploited to determine the color factor ratio
$\caa/\cff$ using the soft particle multiplicities
in gluon and quark jets~\cite{bib-opal99}
(see also~\cite{bib-delphi99}).
}
In the present study,
we present a test of higher order perturbative
calculations based on the total multiplicity
in jets.

The perturbative solutions for multiplicity
are usually expressed
in terms of the anomalous dimension $\gamma$
and the multiplicity ratio~$\ratio$,
defined by
\begin{equation}
  \gamma(y) \equiv \frac{{\ngprime}(y)}{{\ng}(y)} 
   = \left[\ln{\ng}(y)\right]^{\prime} \;\;\;\;\; ; \;\;\;\;\;
  \ratio(y) \equiv \frac{{\ng}(y)}{{\nf}(y)} \;\;\;\;,
  \label{eq-define}
\end{equation}
where $\ng$ and $\nf$ are the mean 
total multiplicities in
gluon and quark jets, respectively.
In these expressions,
$y$=$\ln\,(p\,\theta/\qzero)$ defines the
energy scale of the jets,
with $p$ the momentum of the parton 
which initiates the jet,
$\theta$ the opening angle 
of the first branching in the jet,
and $\qzero$ a cutoff 
which defines the limit of perturbative evolution.
For $p$ large and $\theta$ small,
the virtuality Q$_{\mathrm{jet}}$ (invariant mass) of the jet
is given by Q$_{\mathrm{jet}}$$\approx$$p\,\theta/2$.
Primed quantities in~(\ref{eq-define}) denote
derivatives with respect to~$y$:
${\ngprime}$=d${\ng}/$d$y$, etc.

Perturbative expressions for $\ng$ and $\nf$
at the next-to-leading order (NLO) and the
next-to-next-to-leading order (NNLO) approximations
of pQCD have been presented in~\cite{bib-webber84} 
and~\cite{bib-dremin94}, respectively.
Here we present the solutions at the
next-to-next-to-next-to-leading order (3NLO)
approximation and compare the results to experiment.

The perturbative expressions for $\gamma$ and $\ratio$ 
at 3NLO are:
\begin{eqnarray}
   \gamma(y) & = & \gamma_0(1-\aone\gamma_0-\atwo\gamma_0^2
          -\athree\gamma_0^3) + {\cal{O}}(\gamma_0^5)
     \label{eq-gamma} \\
   \ratio(y) & = & \rzero(1-\rone\gamma_0-\rtwo\gamma_0^2
          -\rthree\gamma_0^3) + {\cal{O}}(\gamma_0^4) \;\;\;\; ,
     \label{eq-ratio}
\end{eqnarray}
with
\begin{equation}
  \gamma_0 = \sqrt{\frac{2\alpha_S{\ncc}}{\pi}}
       = \sqrt{\frac{4{\ncc}}{\beta_0 y}
       \left(1-\frac{\beta_1\ln(2y)}{\beta_0^2 y}\right)}
       \;\;\;\; ,
   \label{eq-gamma0}
\end{equation}
where
$\alpha_S$ is the strong coupling strength,
$\beta_0$=$(11\ncc-2\nff)/3$,
$\beta_1$=$(51\ncc-19\nff)/3$,
$\ncc$=3 is the number of colors,
$\nff$ is the number of active quark flavors,
$\rzero$=$\ncc/\cff$,
and $\cff$=4/3.
The coefficients $a_i$ and $r_i$ are calculable in
pQCD and are given in~\cite{bib-cdnt99} for 
$i$=1,2,3.\footnote{For $\nff$=3,
the value we emphasize in this letter,
$\aone$, $\atwo$ and $\athree$ are 0.280, -0.379 and 0.209,
while $\rone$, $\rtwo$ and $\rthree$ are
0.185, 0.426 and 0.189~\cite{bib-cdnt99}.}

The expressions for $\gamma(y)$
in~(\ref{eq-define}) 
and~(\ref{eq-gamma}) can be combined to yield
\begin{equation}
  {\ng}(y)\propto \exp\left[ \int^{y}
   \,(\gamma _0-\aone\gamma _0^2
    -\atwo\gamma _0^3-\athree\gamma_0^4)\,{\mathrm{d}}y
   \right]
     \;\;\;\; .
\end{equation}
In conjunction with~(\ref{eq-gamma0}),
we then determine
the energy dependence of the gluon jet multiplicity
at 3NLO to be:
\begin{equation}
  {\ng}(y) = 
   {\mathrm{K}}y^{-{\aone}{\cc}^2}\exp{\left[2{\cc}\sqrt{y}
       + \delta_{\mathrm{G}}(y)
\right] }
       \;\;\;\; ,
   \label{eq-ng}
\end{equation}
where K is an overall normalization constant,
${\cc}$=$\sqrt{4{\ncc}/\beta_0}$,
and
\begin{eqnarray}
  \delta_{\mathrm{G}}(y) & = &
    \frac{\cc}{\sqrt{y}} \left\{2{\atwo}{\cc}^2
    +\frac{\beta_1}{\beta_0^2}
      \left[\ln(2y)+2\right]\right\} \nonumber \\
  & &  + \frac{{\cc}^2}{y} \left\{{\athree}{\cc}^2 
    -\frac{{\aone}\beta_1}{\beta_0^2}
    \left[\ln(2y)+1\right]\right\}
  \label{eq-3nlo} \;\;\;\; .
\end{eqnarray}
To obtain this result,
the term proportional to
$\beta_1$ in parenthesis in equation~(\ref{eq-gamma0})
is treated as being small compared to~1.
The first term in the exponent in equation~(\ref{eq-ng})
is the leading order (LO) term.
The term in front of the exponential
arises at NLO.
The term proportional to $1/\sqrt{y}$ in~(\ref{eq-3nlo})
enters at NNLO,
while that proportional to $1/y$ is the 3NLO contribution.

\begin{figure}[p]
\vspace*{-1.5cm}
\begin{center}
\epsfxsize=6 truein
\epsffile[85 100 615 700]{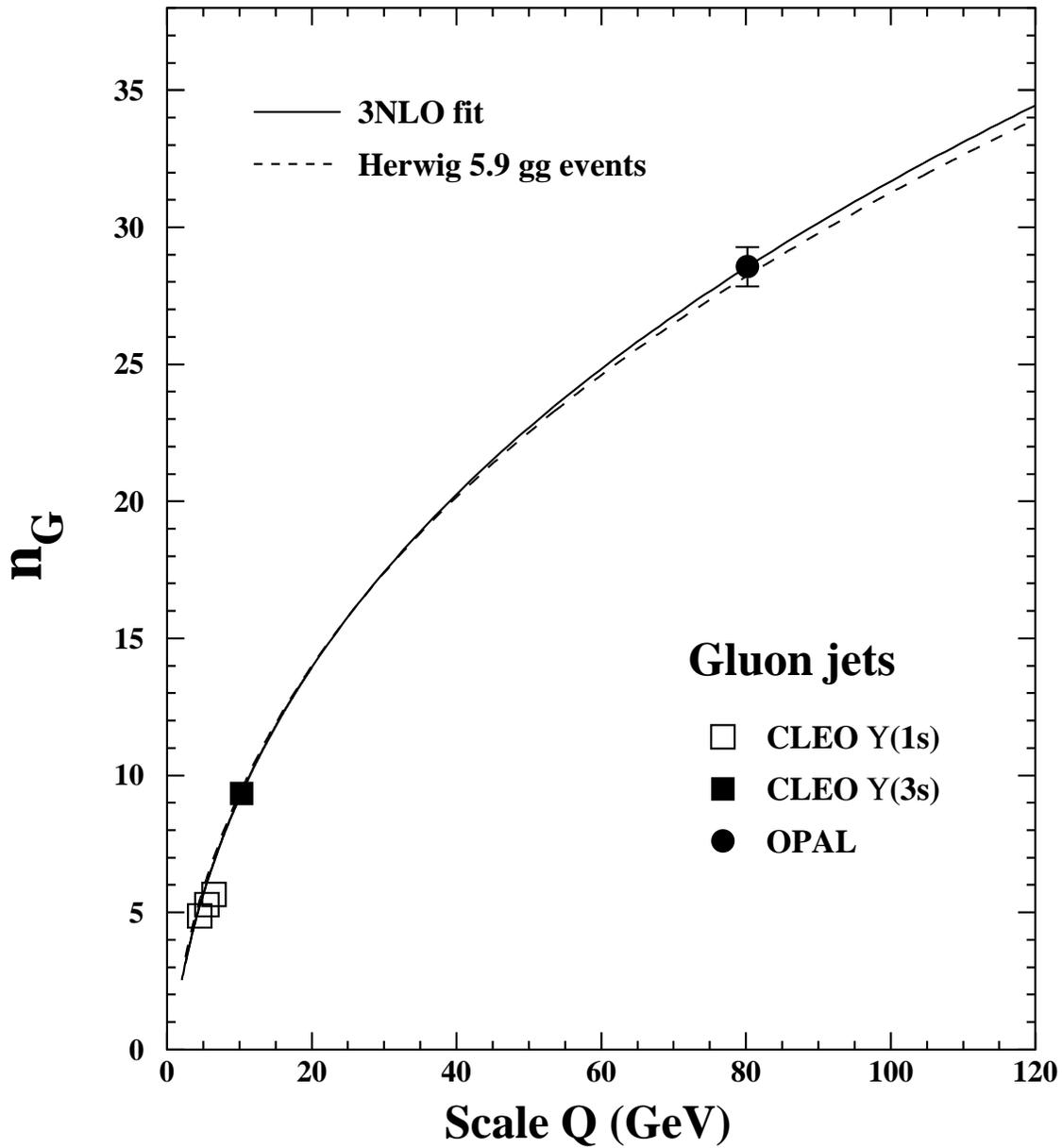}
\end{center}
\vspace*{-1.4cm}
\caption{The mean charged particle multiplicity of 
{\gluglu} events from a color singlet point source
versus energy scale~Q.
The solid curve shows a fit of the 3NLO expression to
the data using $\nff$=3,
where $\nff$ is the number of active quark flavors.
The dashed curve shows the prediction of the Herwig
Monte Carlo for {\gluglu} events.
}
\label{fig-gluons}
\end{figure}

Experimental measurements of the inclusive charged
particle multiplicity of gluon jets
as a function of energy scale Q
are shown in Fig.~\ref{fig-gluons}.
These results utilize a definition
of gluon jets which corresponds to that
employed for the calculation,
based on the production of a virtual gluon jet
pair, {\gluglu},
from a color singlet point source.
These are the only such gluon jet
data currently available.
The three data points at scale Q$\,\approx\,$5~GeV are
derived from the hadronic component of
$\Upsilon$(1S)$\rightarrow\gamma${\gluglu}
decays~\cite{bib-cleo97}.
The virtuality Q is given by the invariant mass of
the hadronic system.
Similarly,
$\Upsilon$(3S)$\rightarrow \gamma\chi_{b2}
$(10.27)$\rightarrow\gamma${\gluglu} 
decays provide the measurement at 
Q$\,\approx\,$10~GeV~\cite{bib-cleo92},
with the scale given by the $\chi_{b2}$ mass.
The result at 
Q$\,\approx\,$80~GeV~\cite{bib-opal99,bib-opal96and98}
is based on hadronic Z$^0$ data:
Z$^0\rightarrow\,$q$\overline{\mathrm{q}}${\gincl},
in which {\gincl} refers to a gluon jet hemisphere
recoiling against two almost collinear quark jets q
and $\overline{\mathrm{q}}$
in the opposite hemisphere.
The {\gincl} jets correspond closely to 
hemispheres of color singlet {\gluglu} events,
as discussed in~\cite{bib-gary94}.
For the results shown here,
the {\gincl} hemisphere
results reported in~\cite{bib-opal99}
have been multiplied by a factor of
two both for the multiplicity and energy scale so that
they correspond to full {\gluglu} events analogous
to the $\Upsilon$ data.

The solid curve in Fig.~\ref{fig-gluons}
shows the result of a $\chi^2$ fit of
expression~(\ref{eq-ng}) to the data,
with $y$=$\ln\,({\mathrm{Q}}/\Lambda)$
where $\Lambda$ is a fitted parameter.
The other fitted parameter is 
the normalization constant~K in~(\ref{eq-ng}).
For this fit, {$\nff$}=3.
The $\Upsilon$(1S) measurements~\cite{bib-cleo97} 
are not included in 
this fit because a systematic uncertainty
was not provided for them.
Nonetheless,
these points are seen to lie near the fitted curve.
The results for the fitted parameters are 
$\Lambda$=$1.03\pm0.24$~GeV and~K=$0.288\pm0.037$.
The uncertainties are defined by the
maximum deviations observed when the
gluon jet measurements are varied by their
one standard deviation uncertainties.
These results,
along with those found using {$\nff$}=4 and~5,
are summarized in the top portion of Table~\ref{tab-nchfit}.
The results of the fits with {$\nff$}=4 and 5 are
virtually indistinguishable from those shown by
the solid curve in Fig.~\ref{fig-gluons}.
Note that the result using {$\nff$}=3 probably has
the most physical relevance since b and c quarks
are rarely produced in the perturbative evolution of the jets,
even for the Z$^0$ data.
The dashed curve in Fig.~\ref{fig-gluons} shows
the prediction of the Herwig parton shower 
multihadronic event 
generator~\cite{bib-herwig}, version~5.9,
for the inclusive charged particle multiplicity
of {\gluglu} events as a function of the
center-of-mass energy,
Q={\ecm}.
The Monte Carlo parameters
are the same as those used in~\cite{bib-opal99}.
Herwig is seen to describe the data well 
and to be similar to the 3NLO result.

The analytic expression for quark jet multiplicity
can be obtained from~(\ref{eq-define}) and~(\ref{eq-ratio}):
\begin{equation}
  \nf(y) = \frac{\ng(y)}{\ratio(y)} =
    \frac{\ng(y)}{
    \rzero\left(1-\rone\gamma_0-\rtwo\gamma_0^2
          -\rthree\gamma_0^3 \right)}
    \;\;\;\; ,
  \label{eq-nq}
\end{equation}
with $\ng(y)$ given by~(\ref{eq-ng}).
The expression for $\ratio(y)$ in the denominator
of~(\ref{eq-nq})
is known to give a relatively poor 
description of data, however.
For example,
the 3NLO prediction at the scale of the Z$^0$
is $\ratio\,$$\approx$$\,1.7$~\cite{bib-cdnt99},
about 13\% larger than the experimental result
of $1.51\pm0.04$~\cite{bib-opal99}.
Therefore,
it can be anticipated that the description
of quark jet multiplicity provided by~(\ref{eq-nq})
will be deficient
if the values of $\Lambda$ and K found using
the gluon jet data are employed.
It should be noted that this problem
with $\ratio$ is even worse
at lower orders,
e.g.~the NLO and NNLO predictions
for $\ratio$ are about $\,2.1$ and 1.8,
respectively,
in even greater disagreement with experiment
than the 3NLO result.

\begin{figure}[p]
\vspace*{-1.5cm}
\begin{center}
\epsfxsize=6 truein
\epsffile[85 100 615 700]{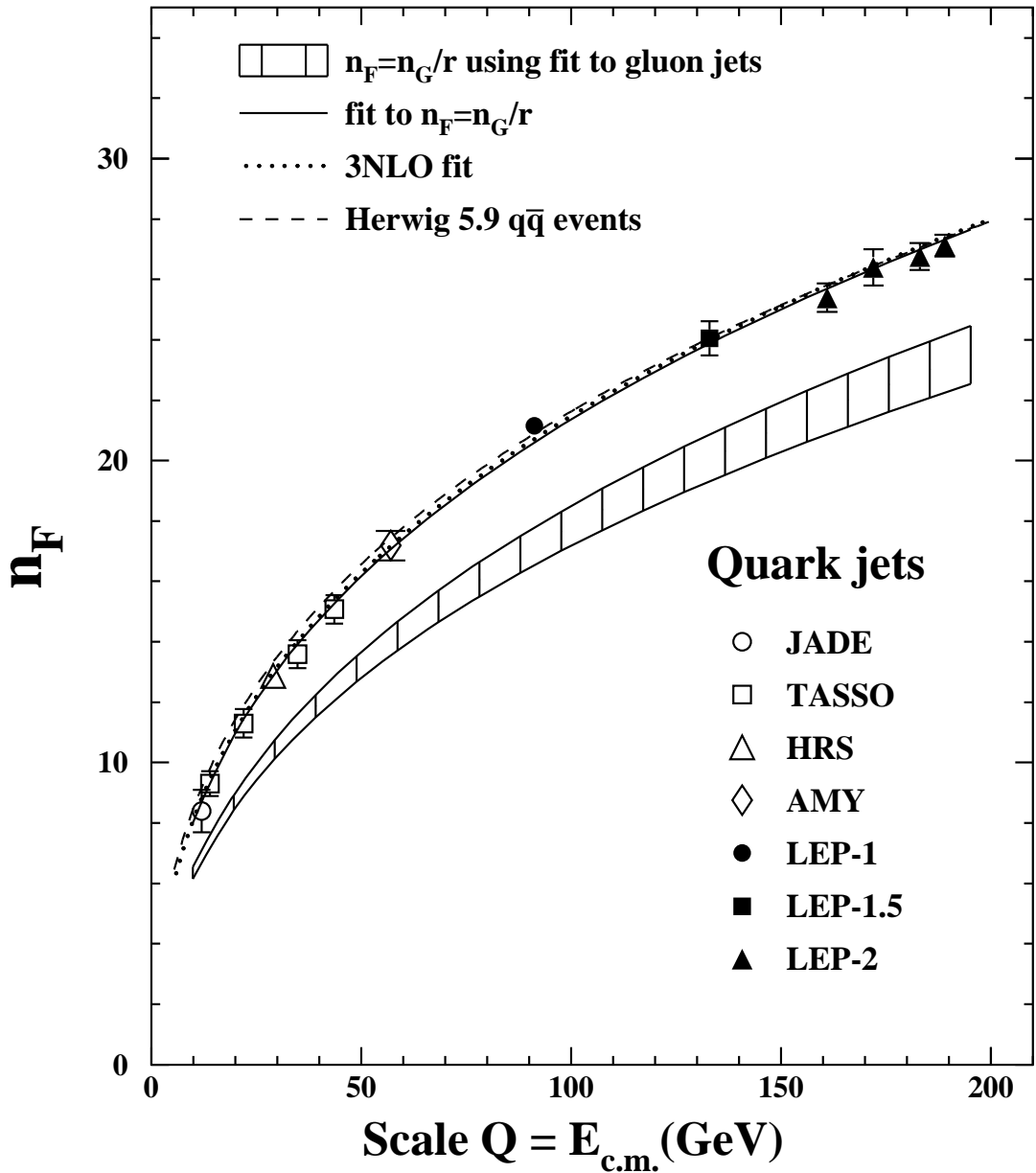}
\end{center}
\vspace*{-1.4cm}
\caption{The mean charged particle multiplicity of {\epem}
hadronic annihilation events versus energy scale~Q={\ecm}.
The dotted curve shows a fit of the 3NLO expression to
the data using $\nff$=3.
The hatched band and solid curve show results using the
expression $\nf$=$\ng/r$ as explained in the text.
The dashed curve shows the prediction of the Herwig
Monte Carlo for {\epem}$\,\rightarrow\,${\qq} events.
}
\label{fig-quarks}
\end{figure}

Experimental measurements of quark jet 
multiplicity are shown in 
Fig.~\ref{fig-quarks} \cite{bib-epemexpts}.
These data are the inclusive charged particle
multiplicity values of
{\epem}$\rightarrow\,($Z$^0/\gamma)^*\rightarrow\,$$hadrons$
events
and correspond to the definition of 
quark jets employed for the calculation,
i.e.~the production of a virtual 
{\qq} pair from a color singlet.
The scale is Q={\ecm}.
The results shown for the LEP experiments
are combined values of ALEPH, DELPHI, L3 and OPAL.
The combined values are obtained using the
unconstrained averaging procedure described in~\cite{bib-pdg98},
for which a common systematic uncertainty is defined
by the unweighted mean of the systematic uncertainties
quoted by the experiments.
LEP-1 refers to data collected at the Z$^0$ peak,
\mbox{LEP-1.5} to data collected at {\ecm}$\approx$133~GeV,
and LEP-2 to data collected at or above the
threshold for W$^+$W$^-$ production.

The vertically striped band in Fig.~\ref{fig-quarks} shows the
prediction of (\ref{eq-nq}) for $\nff$=3
using the values of $\Lambda$ and K from
the fit to the gluon jet measurements 
(Fig.~\ref{fig-gluons}).
The width of the band corresponds to the uncertainties
in the values of $\Lambda$ and K presented above.
Almost identical results to those shown by the band
are obtained using $\nff$=4 or~5.
The band typically lies 15-20\% below 
the data mostly as a consequence of
the problem with the theoretical prediction
for $\ratio$ noted above.

It is also of interest to {\it fit}
expression~(\ref{eq-nq}) to the quark jet measurements.
The results of such a fit using $\nff$=3
are shown by the solid curve in Fig.~\ref{fig-quarks}.
The values obtained for $\Lambda$ and K are
$0.35\pm0.12$~GeV and $0.222\pm0.030$, respectively,
where the uncertainties are defined by
the maximum difference between the results of the
standard fit and those found by
fitting only data between 29 and 189~GeV,
between 12 and 161~GeV,
or by excluding the LEP-1 data point.
The fit is seen to
yield a good description of the data.
The results found using $\nff$=4 or~5 
are essentially identical to those shown by the
solid curve in Fig.~\ref{fig-quarks}.
The values of the fitted parameters are
summarized in the central portion of
Table~\ref{tab-nchfit}.
The values of $\Lambda$ obtained from fitting the
quark jet data are seen to be three to four times
smaller than those obtained from fitting the
gluon jet data.

The dashed curve in Fig.~\ref{fig-quarks} shows
the prediction of Herwig for the inclusive charged particle
multiplicity of hadronic events in
{\epem} annihilations.
The Herwig prediction reproduces the measurements well,
analogously to what was observed for gluon jets.

The expression for $\nf$ given by equation~(\ref{eq-nq}) 
is not entirely satisfactory from a theoretical perspective.
The purely perturbative 3NLO result for $\nf$,
analogous to equation~(\ref{eq-ng}) for $\ng$,
is
\begin{equation}
  {\nf}(y) =
    \frac{\mathrm{K}}{\rzero}\,
    y^{-{\aone}{\cc}^2}\exp{\left[2{\cc}\sqrt{y}
       + \delta_{\mathrm{F}}(y) \right] }
  \label{eq-3nlo-nf}       \;\;\;\; ,
\end{equation}
with
\begin{eqnarray}
  \delta_{\mathrm{F}}(y) & = &
    \frac{\mathrm{C}}{\sqrt{y}} \left\{r_1 + 2{a_2}{\mathrm{C}}^2
    +\frac{\beta_1}{\beta_0^2}
      \left[\ln(2y)+2\right]\right\} \nonumber \\
  & &  + \frac{{\mathrm{C}}^2}{y} 
    \left\{  a_3{\mathrm{C}}^2 + \frac{r_1^2}{2} + r_2
    -\frac{{a_1}\beta_1}{\beta_0^2}
    \left[\ln(2y)+1\right]\right\} 
  \label{eq-delf} \;\;\;\; .
\end{eqnarray}
where the same notation is used as in
equations~(\ref{eq-ng}) and~(\ref{eq-3nlo}).
Equation~(\ref{eq-3nlo-nf}) is derived from
equation~(\ref{eq-nq}),
keeping terms only up to $y^{-1}$ in the exponent.
It is seen that a term proportional 
to $\rthree$ appears in~(\ref{eq-nq}) 
but not in~(\ref{eq-3nlo-nf}):
such a term would be proportional to $y^{-3/2}$
in~(\ref{eq-delf}) and thus would contribute
to the 4NLO approximation.
This difference between~(\ref{eq-nq}) and~(\ref{eq-3nlo-nf})
arises because of the difference in the order of $\gamma_0$  
in~(\ref{eq-gamma}) and~(\ref{eq-ratio}).
The manner in which the powers of $\gamma_0$
are structured in~(\ref{eq-gamma}) 
and~(\ref{eq-ratio}) is conventional 
because $\gamma$ enters in the exponent whereas 
$\ratio$ is used as a multiplicative factor. 
As a consequence of this difference
between~(\ref{eq-gamma}) and~(\ref{eq-ratio}),
all corrections of a given order in the denominator
of~(\ref{eq-nq}) contribute
to yet higher orders in~(\ref{eq-3nlo-nf}).
For example, 
$\rone$ contributes only to the 
NLO term in the denominator of~(\ref{eq-nq})
whereas it contributes to the 
NNLO ($\sim y^{-1/2}$)and 3NLO ($\sim y^{-1}$)
terms in~(\ref{eq-3nlo-nf}).
This situation can also be considered to arise from
the fact that the logarithmic slope of quark jets,
$\gamma_{\mathrm{F}}$, is given by
\begin{equation}
  \gamma_{\mathrm{F}} \equiv
     \left[\ln{\nf}\right]^{\prime}
   = \gamma - \frac{\rprime}{\ratio}
  \label{eq-gammaf} \;\;\;\; .
\end{equation}
However,
from~(\ref{eq-ratio}) and~(\ref{eq-gamma0}),
$\rprime\approx-\rzero\,\rone\gamma_0^{\prime}
\sim\rzero\,\rone\gamma_0^3$.
Therefore,
the first correction to $\gamma_{\mathrm{F}}$
appears at NNLO,
but it is determined by the NLO correction to~$\ratio$.
We conclude that
it is improper to use the term with
$\rthree$ in~(\ref{eq-nq}) until the 4NLO
contribution to $\gamma$ is known.
By extension
this implies that if the NLO formula is used
to describe quark jet multiplicity,
as is common practice (see below),
the LO result $\ratio$=$\rzero$ should be
inserted into all formulas
to be self-consistent within the perturbative approach.
Therefore the mean multiplicities in quark and gluon jets 
coincide in NLO up
to the constant normalization factor $\rzero$=9/4. 
Any usage of $\ratio$ different from $\rzero$ implies
that NNLO corrections have been included.

The dotted curve in Fig.~\ref{fig-quarks}
shows the result of a $\chi^2$ fit of
equation~(\ref{eq-3nlo-nf}) to the quark jet data
for $\nff$=3.
The latter equation is seen to provide a
good description of the measurements,
equivalent to that given by equation~(\ref{eq-nq}).
The fitted curves for $\nff$=4 and~5 are
indistinguishable from the curve for $\nff$=3.
The values of the fitted parameters are given
in the bottom section of Table~\ref{tab-nchfit}.
These values are similar to those found from
fitting equation~(\ref{eq-nq}) to the
quark jet data.

Our method of comparing the 3NLO predictions 
to data is in contrast to the procedure usually
employed to test QCD analytic
predictions for multiplicity.
Typically,
the NLO expression for gluon jets~\cite{bib-webber84}
is used to fit quark jet data.
A fit to gluon jet data is rarely performed.
At NLO,
the theoretical expressions for $\nf$ and $\ng$
differ by only the constant factor~$\rzero$,
as noted above,
which justifies this procedure.
The NLO expression is found to provide a good
description of the measurements,
analogous to the agreement we find in the
fit of equations~(\ref{eq-nq}) 
or~(\ref{eq-3nlo-nf}) to quark jets
(the solid or dotted curves in Fig.~\ref{fig-quarks}).
This good description belies the
fact that the appropriate
prediction for $\ratio$ at this order, $\rzero$=2.25,
is in striking disagreement with the
experimental value $\ratio$$\approx$1.5.

\begin{figure}[p]
\vspace*{-1.5cm}
\begin{center}
\epsfxsize=6 truein
\epsffile[85 100 615 700]{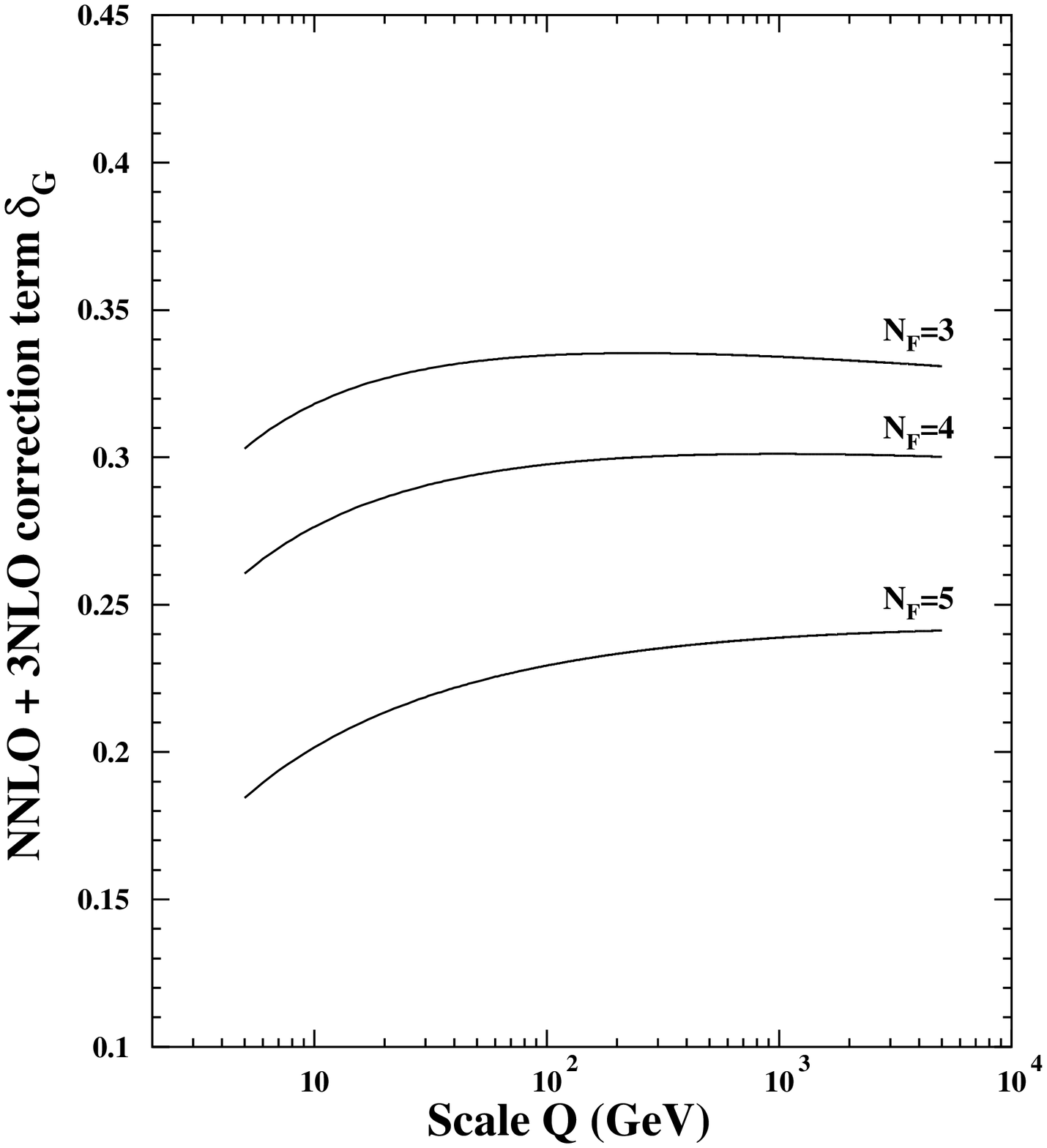}
\end{center}
\vspace*{-1.4cm}
\caption{The behavior of the sum of the NNLO and 3NLO
correction term to the gluon jet multiplicity,
$\delta_{\mathrm{G}}$,
as a function of energy scale~Q,
for $\nff$=3, 4 and~5.
}
\label{fig-del}
\end{figure}

The good description of quark jet multiplicity at NLO
is easily understood,
because the sum of the NNLO 
and 3NLO corrections to~$\ng$,
given by~$\delta_{\mathrm{G}}(y)$~(equation~(\ref{eq-3nlo})),
is small and almost constant at presently accessible
energies.
To illustrate this,
we show in Fig.~\ref{fig-del}
the behavior of $\delta_{\mathrm{G}}(y)$ as a function
of scale for $\nff$=3, 4 and~5.
For the results of this figure,
the scale is $y$=$\ln(\mathrm{Q}/\Lambda)$ for which
the values of $\Lambda$ are taken from the fits to quark
jet data using equation~(\ref{eq-nq})
(central section of Table~\ref{tab-nchfit})
to correspond to the procedure outlined above
to test the NLO equation.
It is seen that $\delta_{\mathrm{G}}$ 
is almost constant for the range of
energies relevant to experiments.
Furthermore, $\delta_{\mathrm{G}}$ is small compared
to the LO term in the exponent of~(\ref{eq-ng}),
which typically has a value from 5 to 7 over
the range of scale Q shown.
Therefore the NLO formula for $\ng$
is not modified significantly by the NNLO and 3NLO
contributions and the expression $\nf$=$\ng/\ratio$
fits the quark jet data well irrespective of whether
the higher order terms are included in~$\ng$.

In summary,
we have presented analytic expressions for
gluon and quark jet multiplicity as a
function of energy scale at the 
next-to-next-to-next-to-leading order (3NLO)
of perturbation theory.
We have performed fits of these expressions to 
gluon and quark jet measurements,
using data for which the experimental and
theoretical definitions of jets coincide.
To our knowledge,
this is the first time such a fit
has been performed for gluon jets.
We find that the gluon jet expression
describes the available gluon jet measurements
accurately,
with values of the parameter $\Lambda$ in the
range from about 0.6 to 1.0~GeV depending
on the number of active quark flavors,~$\nff$.
The 3NLO expression for quark jets also
provides an accurate description of the data
with values of $\Lambda$
about three times smaller than those found
for gluon jets.
We attribute the well known agreement of the
next-to-leading order (NLO) expression for 
quark jet multiplicity with data to the facts that
the gluon and quark jet predictions are
equivalent at this order to within the constant
factor $\rzero$=9/4,
and that the gluon jet
prediction is only slightly modified by the
NNLO and 3NLO terms.

\newpage

\begin{table}[t]
\centering
\begin{tabular}{|c|ccc|}
 \hline
  & & &  \\[-2.4mm]
  & $\Lambda$ (GeV) & K & $\chi^2$/d.o.f.  \\[2mm]
 \hline
 \hline
 \multicolumn{4}{|l|}{}  \\[-2.4mm]
 \multicolumn{4}{|l|}{ (a) Gluon jets,
               equation~(\ref{eq-ng})} \\[2mm]
 \hline
  & & &  \\[-2.4mm]
$\nff$=3 & $1.03\pm0.24$  &  $0.288\pm0.037$ & 0.01/2\\[2mm]
$\nff$=4 & $0.84\pm0.21$  &  $0.244\pm0.034$ & 0.01/2\\[2mm]
$\nff$=5 & $0.64\pm0.17$  &  $0.205\pm0.031$ & 0.01/2\\[2mm]
 \hline
 \multicolumn{4}{|l|}{}  \\[-2.4mm]
 \multicolumn{4}{|l|}{ (b) Quark jets,
               equation~(\ref{eq-nq})} \\[2mm]
 \hline
  & & &  \\[-2.4mm]
$\nff$=3 & $0.35\pm0.12$    &  $0.222\pm0.030$ & 5.8/13\\[2mm]
$\nff$=4 & $0.233\pm0.086$  &  $0.170\pm0.024$ & 5.9/13\\[2mm]
$\nff$=5 & $0.135\pm0.055$  &  $0.126\pm0.019$ & 6.1/13\\[2mm]
 \hline
 \multicolumn{4}{|l|}{}  \\[-2.4mm]
 \multicolumn{4}{|l|}{ (b) Quark jets,
               equation~(\ref{eq-3nlo-nf})} \\[2mm]
 \hline
  & & &  \\[-2.4mm]
$\nff$=3 & $0.322\pm0.098$  &  $0.216\pm0.025$ & 5.6/13\\[2mm]
$\nff$=4 & $0.231\pm0.083$  &  $0.169\pm0.024$ & 5.8/13\\[2mm]
$\nff$=5 & $0.148\pm0.059$  &  $0.129\pm0.020$ & 6.1/13\\[2mm]
 \hline
\end{tabular}
\caption{
Results of a two parameter fit of QCD expressions 
for the scale evolution of event multiplicity 
to the measured mean charged particle multiplicities of
(a)~{\gluglu} (top),
and (b) and (c)~{\qq} events (center and bottom)
from a color singlet point source.
For the gluon jets,
the QCD expression used is the exact 3NLO result
for $\ng$, equation~(\ref{eq-ng}).
For the quark jets,
the QCD expression is given either by
(b)~$\nf$=$\ng/\ratio$, equation~(\ref{eq-nq}),
which contains some terms beyond 3NLO as
discussed in the text,
or by~(c) the exact 3NLO result for $\nf$,
equation~(\ref{eq-3nlo-nf}).
$\Lambda$ is an energy scale parameter
related to the perturbative cutoff~$\qzero$
while K~is an overall normalization constant.
For the gluon jets,
the uncertainties of $\Lambda$ and K
are evaluated by varying the
gluon jet data points by $\pm~1$~standard deviation
of their total uncertainties.
For the quark jets,
the uncertainties are evaluated
by varying the fit range as described in the text.
}
\label{tab-nchfit}
\end{table}

\newpage

\newpage

\newpage

\newpage

\begin{thebibliography}{99}
\bibitem{bib-reviews}
  See, for example, I.M. Dremin, 
  Physics-Uspekhi {\bf 37} (1994)~715; \newline
  V.A. Khoze and W. Ochs, 
  Int. J. Mod. Phys. {\bf A12} (1997)~2949,\newline
  and references therein.
\bibitem{bib-opal99}
  OPAL Collaboration, G. Abbiendi {\it et al.},
  CERN-EP/99-028,
  in press in Eur. Phys. J. {\bf C}.
\bibitem{bib-delphi99}
  DELPHI Collaboration, P. Abreu {\it et al.},
  Phys. Lett. {\bf B449} (1999)~383.
\bibitem{bib-webber84}
  B.R. Webber, Phys. Lett. {\bf B143} (1984)~501.
\bibitem{bib-dremin94}
  I.M. Dremin and V.A. Nechitailo,
  Mod. Phys. Lett. {\bf A9} (1994)~1471.
\bibitem{bib-cdnt99}
  A Capella, I.M. Dremin, J.W. Gary,
  V.A. Nechitailo, J. Tran Thanh Van,
  submitted to Nucl. Phys.~{\bf B}.
\bibitem{bib-cleo97}
  CLEO Collaboration, M.S. Alam {\it et al.},
  Phys. Rev. {\bf D56} (1997)~17.
\bibitem{bib-cleo92}
  CLEO Collaboration, M.S. Alam {\it et al.},
  Phys. Rev. {\bf D46} (1992)~4822.
\bibitem{bib-opal96and98}
  OPAL Collaboration, G. Alexander {\it et al.},
  Phys. Lett. {\bf B388} (1996)~659; \newline
  OPAL Collaboration, K. Ackerstaff {\it et al.},
  Eur. Phys. J. {\bf C1} (1998)~479. 
\bibitem{bib-gary94}
  J.W. Gary, Phys. Rev. {\bf D49} 4503~(1994).
\bibitem{bib-herwig}
  G. Marchesini, B.R. Webber {\it et al.},
  Comp. Phys. Comm. {\bf 67} 465~(1992).
\bibitem{bib-epemexpts}
  JADE Collaboration, W. Bartel {\it et al.}, 
  Z. Phys. {\bf C20} (1983)~187;\newline
  TASSO Collaboration, W. Braunschweig {\it et al.}, 
  Z. Phys. {\bf C45} (1989)~193;\newline
  HRS Collaboration, M. Derrick {\it et al.}, 
  Phys. Rev. {\bf D34} (1986)~3304;\newline
  AMY Collaboration, H.W. Zheng {\it et al.}, 
  Phys. Rev. {\bf D42} (1990)~737;\newline
  ALEPH note 98-014 (contribution to the summer conferences)
  ;\newline
  ALEPH LEPC presentation, November 1998;\newline
  DELPHI Collaboration, P. Abreu {\it et al.},
  Eur. Phys. J. {\bf C6} (1999)~19;\newline
  DELPHI Collaboration, P. Abreu {\it et al.},
  Phys. Lett. {\bf B372} (1996)~172;\newline
  DELPHI Collaboration, P. Abreu {\it et al.},
  Phys. Lett. {\bf B416} (1998)~233;\newline
  L3 Collaboration, B. Adeva {\it et al.},
  Z. Phys. {\bf C55} (1992)~39;\newline
  L3 Collaboration, M. Acciarri {\it et al.},
  Phys. Lett. {\bf B371} (1996)~137;\newline
  L3 Collaboration, M. Acciarri {\it et al.},
  Phys. Lett. {\bf B404} (1997)~390;\newline
  L3 Collaboration, M. Acciarri {\it et al.},
  Phys. Lett. {\bf B444} (1998)~569;\newline
  OPAL Collaboration, K. Ackerstaff {\it et al.},
  CERN-EP/98-089;\newline
  OPAL Collaboration, G. Alexander {\it et al.},
  Z. Phys. {\bf C72} (1996)~191;\newline
  OPAL Collaboration, K. Ackerstaff {\it et al.},
  Z. Phys. {\bf C75} (1997)~193;\newline
  OPAL Physics Note PN281 (Feb. 1997);\newline
  OPAL Physics Note PN323 (Nov. 1997);\newline
  OPAL LEPC presentation, November 1998.
\bibitem{bib-pdg98}
  Review of Particle Physics,
  Eur. Phys. J. {\bf C3} (1998)~1.
\end{thebibliography}
\end{document}